\documentclass[reqno]{amsart}

\usepackage{booktabs}
\usepackage{IEEEtrantools}
\usepackage{amssymb,latexsym,amsfonts,amsmath}
\usepackage{mathrsfs}
\usepackage{graphicx}
\usepackage{dsfont}
\usepackage{enumitem}

\usepackage{mdwlist}
\usepackage{refcount}
\usepackage{comment}

\usepackage{tikz}
\usetikzlibrary{calc,shapes,arrows}

\usepackage{algorithm}
\usepackage{algorithmic}

\usepackage{xspace}

\newtheorem{theorem}{Theorem}[section]
\newtheorem{lemma}[theorem]{Lemma}

\newtheorem{definition}[theorem]{Definition}

\newtheorem{remark}[theorem]{Remark}
\newtheorem{assumption}{Assumption}
\numberwithin{equation}{section}

\newcommand{\R}{{\mathbb{R}}}

\newcommand{\Let}{:=}
\newcommand{\EE}{\mathds{E}}
\newcommand{\PP}{\mathds{P}}

\usepackage{fancyhdr}

\newenvironment{nouppercase}{%
	\renewcommand{\uppercasenonmath}[1]{}}{}

\begin{document}

\begin{abstract}
This work is concerned with the safety controller synthesis of stochastic hybrid systems, in which continuous evolutions are described by stochastic differential equations with both Brownian motions and Poisson processes, and instantaneous jumps are governed by stochastic
difference equations with additive noises. Our proposed framework leverages the notion of \emph{control barrier certificates} (CBC), as a \emph{discretization-free} approach, to synthesize safety controllers for stochastic hybrid systems while providing safety guarantees in finite time horizons. In our proposed scheme, we first provide an augmented framework to characterize each stochastic hybrid system containing continuous evolutions and instantaneous jumps with a unified system covering both scenarios. We then introduce an augmented control barrier certificate (ACBC) for augmented systems and propose sufficient conditions to construct an ACBC based on CBC of original hybrid systems. By utilizing the constructed ACBC, we quantify upper bounds on the probability that the stochastic hybrid system reaches certain unsafe regions in a finite time horizon. The proposed approach is verified over a nonlinear case study.
\end{abstract}

\title{{\LARGE Safety Barrier Certificates for Stochastic Hybrid Systems}$^*$\footnote[1]{$^*$This work was supported in part by the Swiss National Science Foundation under NCCR Automation, grant agreement 51NF40-180545, and by the UKRI EPSRC New Investigator Award CodeCPS (EP/V043676/1).}}

\author{{\bf {\large Abolfazl Lavaei}}$^1$}
\author{{\bf {\large Sadegh Soudjani}}$^2$}
\author{{\bf {\large Emilio Frazzoli}}$^1$\\
	{\normalfont $^1$Institute for Dynamic Systems and Control, ETH Zurich, Switzerland}\\
	{\normalfont $^2$School of Computing, Newcastle University, United Kingdom}\\
\texttt{\{alavaei,efrazzoli\}@ethz.ch}, \texttt{Sadegh.Soudjani@ncl.ac.uk}}

\pagestyle{fancy}
\lhead{}
\rhead{}
  \fancyhead[OL]{Abolfazl Lavaei, Sadegh Soudjani, Emilio Frazzoli}

  \fancyhead[EL]{Safety Barrier Certificates for Stochastic Hybrid Systems} 
  \rhead{\thepage}
 \cfoot{}
 
\begin{nouppercase}
	\maketitle
\end{nouppercase}

\section{Introduction}

This work is motivated by the challenges arising in the controller synthesis of continuous-space stochastic hybrid systems (SHSs). Over the past two decades, SHSs have become ubiquitous as a pivotal modeling framework playing significant roles in many safety-critical applications. Models of SHSs are inherently heterogeneous: discrete systems describe computational parts and continuous dynamics characterize physical processes. In addition, SHSs can contain both continuous evolution and instantaneous jumps. Accordingly, the ability to handle the interaction between continuous and discrete dynamics (in both space and time) is a prerequisite for acquiring a rigorous formal framework for verification and synthesis of SHSs. 

Since the complexity raised by the aforesaid interaction often prevents one to acquire analytical solutions, the verification and controller synthesis of SHSs are often addressed by methods of (in)finite abstractions (\cite{julius2009approximations,APLS08,zamani2015symbolic,zamani2014symbolic,tmka2013}). More concretely, since the closed-form solution of synthesized controllers for SHSs is not available in general, a promising approach is to approximate original models by a simpler one with either a lower dimension (\emph{a.k.a.,} infinite abstractions) or with discrete-state sets  (\emph{a.k.a.,} finite abstractions). However, the proposed abstraction-based techniques hinge on the discretization of state and input sets, and consequently, they suffer severely from the \emph{curse of dimensionality} problem: the complexity exponentially grows  with the dimension of the system. Hence, \emph{compositional} abstraction-based techniques have been proposed in the past few years to construct abstractions of complex SHSs based on abstractions of smaller subsystems (\cite{SAM17,hahn2013compositional,lavaei2018CDCJ,lavaei2019HSCC_J,lavaei2018ADHSJ,Lavaei_TAC2022,Lavaei_Survey,AmyJournal2020}).

Another promising approach, proposed in the past decade, for the formal verification and controller synthesis of complex dynamical systems is to employ \emph{control barrier certificates} as a \emph{discretization-free} technique. This approach is initially proposed for formal analysis of (stochastic) hybrid systems in~\cite{prajna2004safety,wieland2007constructive,prajna2007framework} and has received significant attentions in the past few years. Intuitively speaking, barrier certificates are Lyapunov-like functions defined over the state space of the system to enforce a set of inequalities on both the function itself and its infinitesimal generator along the flow (or one-step transition) of the system. A suitable level set of a barrier certificate separates an unsafe region from all system trajectories starting from a given set of initial states. As a result, the existence of such a function provides a formal (probabilistic) certificate for the safety of the system. Barrier certificates have been so far widely employed for formal verification and synthesis of non-stochastic \cite{borrmann2015control,wang2017safety,ames2019control} and stochastic dynamical systems \cite{zhang2010safety,yang2020efficient,M.Ahmadi,ahmadi2019safe,santoyo2019verification,clark2019control,LavaeiIFAC2020,AmyAutomatica2020_J,Niloofar_TNCS_2022,Amy_LCSS20}, to name a few.

Although existing results on the formal analysis of dynamical systems via barrier certificates are comprehensive, unfortunately, there exist no results on the safety controller synthesis of SHSs with both continuous evolutions and instantaneous jumps. Our main contribution here is to propose, for the first time, a construction scheme for the formal controller synthesis of SHSs, in which underlying dynamics contain both continuous evolutions modeled by stochastic differential equations with Brownian motions and Poisson processes, and instantaneous jumps governed by stochastic difference equations with additive noises. To do so, we first propose an augmented framework to describe each SHS containing continuous evolutions and instantaneous jumps with a unified system covering both scenarios, whose state trajectories are exactly the same as those of original hybrid systems. We then introduce an augmented control barrier certificate (ACBC) for augmented systems by proposing required conditions for the construction of ACBC based on CBC of original hybrid systems. We leverage the constructed ACBC and quantify upper bounds on the probability that the SHS reaches certain unsafe regions in finite time horizons. Proofs of most statements are omitted due to space limitations.

Construction of symbolic models for a class of impulsive systems is presented in~\cite{swikir2020symbolic}. Our approach here differs from the one in~\cite{swikir2020symbolic} in two main directions. First and foremost, our results are based on control barrier certificates, as a \emph{discretization-free} approach, whereas the abstraction-based technique proposed in~\cite{swikir2020symbolic} relies on the discretization of state and input sets, and consequently, it suffers severely from the \emph{curse of dimensionality} problem. Second, our proposed framework here deals with \emph{stochastic hybrid systems} in which continuous evolutions are characterized by stochastic differential equations with Brownian motions and Poisson processes, and instantaneous jumps are governed by stochastic difference equations with additive noises. In contrast, the results in~\cite{swikir2020symbolic} only handle non-stochastic impulsive systems.

\section{Stochastic Hybrid Systems}\label{Sec:SHSs}

\subsection{Notation and Preliminaries}

The following notation is employed throughout the paper. We denote sets of real, positive and non-negative real numbers by $\mathbb{R},\mathbb{R}_{>0}$, and $\mathbb{R}_{\geq 0}$, respectively. We use $\mathbb{R}^n$ to denote a real space of $n$ dimension. $\mathbb{N} := \{0,1,2,...\}$ represents the set of non-negative integers and $\mathbb{N}_{\geq 1}=\{1,2,...\}$ is the set of positive integers. Given $N$ vectors $x_i \in \mathbb{R}^{n_i}$, $x=[x_1;...;x_N]$ denotes the corresponding vector of dimension $\sum_i n_i$. Given a matrix $A\in\R^{N\times{N}}$ with diagonal entries $a_1,\ldots,a_N$, we define $\textsf{Tr}(A) = \sum_{i=1}^N a_i$. Given a measurable function $f:\mathbb N\rightarrow\mathbb{R}^n$, the (essential) supremum of $f$ is denoted by $\Vert f\Vert_{\infty} \Let \text{(ess)sup}\{\Vert f(k)\Vert,k\geq 0\}$.

We consider a probability space $(\Omega,\mathcal F_{\Omega},\mathds{P}_{\Omega})$, where $\Omega$ is the sample space,
$\mathcal F_{\Omega}$ is a sigma-algebra on $\Omega$ comprising subsets of $\Omega$ as events, and $\mathds{P}_{\Omega}$ is a probability measure that assigns probabilities to events. We assume that triple $(\Omega,\mathcal F_{\Omega},\mathds{P}_{\Omega})$ is endowed with a filtration $\mathbb{F} = (\mathcal F_s)_{s\geq 0}$ satisfying the usual conditions of completeness and right continuity. Let $(\mathbb W_s)_{s \ge 0}$ be a ${\textsf b}$-dimensional $\mathbb{F}$-Brownian motion, and $(\mathbb P_s)_{s \ge 0}$ be an ${\textsf r}$-dimensional $\mathbb{F}$-Poisson process. We assume that the Poisson process and Brownian motion are independent of each other. The Poisson process $\mathbb P_s = [\mathbb P_s^1; \cdots; \mathbb P_s^{\textsf r}]$ models ${\textsf r}$ events whose occurrences are assumed to be independent of each other.

\subsection{Stochastic Hybrid Systems}
In this work, we study stochastic hybrid systems (SHSs) with both continuous evolutions and instantaneous jumps, in which continuous evolutions are modeled by stochastic differential equations with Brownian motions and Poisson processes, and instantaneous jumps are governed by stochastic
difference equations with additive noises. We formalize this class of SHSs in the following definition.

\begin{definition}\label{Def:SHS}
	A stochastic hybrid system (SHS) $\Sigma$ is defined by the tuple	$\Sigma=(\mathbb{R}^n, U,\mathcal{U},\sigma,\rho,f_1,\varsigma,f_2)$,
	where: 
	\begin{itemize}
		\item $\mathbb{R}^n$ is the state space of the
		system;
		\item $U\subseteq\mathbb{R}^m$ is the input space of the system;
		\item $\mathcal U$ is the set of all measurable bounded input functions $\nu:\mathbb{R}_{\geq0}\rightarrow  U$;
		\item  $\sigma: \mathbb R^n \rightarrow \mathbb R^{n\times \textsf b}$ is the diffusion term which is globally Lipschitz continuous;
		\item $\rho: \mathbb R^n \rightarrow \mathbb R^{n\times \textsf r}$ is the reset term which is globally Lipschitz continuous;
		\item $f_1: \mathbb{R}^n\times  U \rightarrow \mathbb{R}^n $ is the drift term which is globally Lipschitz continuous;
		\item $\varsigma$ is a sequence of independent and identically distributed
		(i.i.d.) random variables from a sample space $\Omega$ to the measurable space $(\mathcal{V}_\varsigma,\mathcal F_\varsigma)$, namely,	 
		\begin{equation*}
		\varsigma:=\Big\{\varsigma(\cdot)\!:(\Omega,\mathcal F_\Omega)\rightarrow (\mathcal{V}_\varsigma,\mathcal F_\varsigma)\Big\}; 
		\end{equation*}
		\item $f_2: \mathbb{R}^n\times U \times \mathcal V_{\varsigma}\rightarrow \mathbb{R}^n $ is the transition map which is globally Lipschitz continuous.
	\end{itemize}
	
	The stochastic hybrid system $\Sigma$ is described by stochastic differential and difference equations of the form
	\begin{align}\label{Eq:1}
	\Sigma\!:\left\{\hspace{-2mm}
	\begin{array}{rl}
	\mathsf{d}x(t)\!=\!\!\!\!& f_1(x(t),\nu(t))\mathsf{d}t+\sigma(x(t))\mathsf{d}\mathbb W_t+\rho(x(t))\mathsf{d}\mathbb P_t,~ t\in\mathbb{R}_{\geq0}\backslash \Lambda,\\
	x(t)\!=\!\!\!\!& f_2(x(t^-),\nu(t),\varsigma(t)),\quad\quad\quad\quad\quad\quad\quad\quad\quad\quad\quad \!\!\!\!\!t\in \Lambda,
	\end{array}
	\right.
	\end{align}
	where $\Lambda=\{t_k\}_{k\in\mathbb N}$ with $t_{k+1}-t_{k}\in\{q_1\tau,\ldots,q_2\tau\}$ for fixed jump parameters $\tau\in\mathbb R_{>0}$ and $q_1,q_2\in \mathbb N_{\ge1}$, $q_1\le q_2$, and $f_2(x(t^-)) = \lim_{t\to t_k^-} f_2(x(t))$, \emph{i.e.,} the left limit of $f_2(x(t))$ when $t$ approaches $t_k$ from left. In addition, $x:\mathbb R_{\geq0}\rightarrow \mathbb R^n $ is the state signal, which is assumed to be right-continuous for all $t\in\mathbb R_{\ge0}$, and $\nu(\cdot)\in\mathcal{U}$ is the input signal. The random sequence $x_{x_0,\nu}(t)$ satisfying~\eqref{Eq:1} for any initial state $x_0=x(0)\in \mathbb R^n$ under an input signal  $\nu(\cdot)\in\mathcal{U}$ at time $t\in \mathbb R_{\geq0}$ is called the \textit{solution process} of $\Sigma$ under the input $\nu$ and the initial state $x_0$. We also denote by $\Sigma_1$ and $\Sigma_2$ the continuous- and discrete-time dynamics of the system $\Sigma$, \emph{i.e.,}
	\begin{align*}
	&\Sigma_1\!\!:\mathsf{d}x(t)= f_1(x(t),\nu(t))\mathsf{d}t+\sigma(x(t))\mathsf{d}\mathbb W_t +\rho(x(t))\mathsf{d}\mathbb P_t,\\
	&\Sigma_2\!\!:x(t)= f_2(x(t^-),\nu(t), \varsigma(t)).
	\end{align*}
\end{definition}	

In this work, we restrict our attention to sampled-data hybrid systems, where input curves belong to $\mathcal{U}_{\tau}$ taking a constant
value for a duration $\tau$, \emph{i.e.,}
\begin{align}\label{Eq:2}
\mathcal{U}_\tau=\Big\{\nu\!:\mathbb R_{\ge0}\rightarrow U\,\big|\, \nu(t)=\nu((k-1)\tau), ~t\in [(k-1)\tau,k\tau)],k\in \mathbb N_{\geq1}\Big\}.
\end{align}
We denote discrete time instances by $t_k = k\tau, k\in\mathbb N$.

\begin{remark}
	Stochastic hybrid systems studied in this work have broad applications in real-life safety-critical systems such as biological networks, communication networks, power grids, health and epidemiology, air traffic networks, and manufacturing systems~\cite{blom2006stochastic,cassandras2018stochastic,hespanha2004stochastic}, to name a few.
\end{remark}
\subsection{Augmented Stochastic Hybrid Systems}

Here, in order to describe each SHS with a unified system covering both continuous evolutions and instantaneous jumps, we provide an alternative characterization of SHS, called augmented stochastic hybrid systems (ASHS), as formalized in the next definition.  

\begin{definition}\label{Def:ASHS}
	Given a stochastic hybrid system $\Sigma=(\mathbb{R}^n, U,\mathcal{U}_\tau,\sigma,\rho,f_1,\varsigma,f_2)$ with jump parameters ($\tau$, $q_1$, $q_2$), we define the associated augmented stochastic hybrid system (ASHS) $\mathbb{A}(\Sigma)=(\mathbb X,\mathbb U,\sigma,\rho,\varsigma, \mathbb F,\mathbb Y,\mathbb H)$, where
	\begin{itemize}
		\item  $\mathbb X=\mathbb R^n \times \{0,\ldots,q_2\}$ is the set of states, in which $(x,z) \in \mathbb{X}$ denotes that the current state of $\Sigma$ is $x$, and the time elapsed since the latest jump capped by $q_2$ is $z$;
		\item $\mathbb U=\mathcal{U}_{\tau}$ is the set of inputs;
		\item $\sigma$ is the diffusion term;
		\item $\rho$ is the reset term;
		\item $\varsigma$ is a sequence of i.i.d. random variables;
		\item  $(x',z')= \mathbb F((x,z),\nu,\sigma,\rho,\varsigma)$ if and only if one of the following two scenarios holds:
		\begin{enumerate}[label=(\roman*)]
			\item Flow scenario: $0\leq z\leq q_{2}-1$, $x'=  x_{x,\nu}(\tau^-)$, and $z'=z+1$;
			\item Jump scenario: $q_1\leq z\leq q_2$,  $x'= f_2(x,\nu,\varsigma)$, and $z'=0$; 
		\end{enumerate}
		\item $\mathbb Y=\mathbb R^n$ is the output space;
		\item $\mathbb H:\mathbb X\rightarrow \mathbb Y$ is the output map defined as $\mathbb H(x,z)=x$.
	\end{itemize}
\end{definition}

\begin{remark}
	Note that, in ASHS $\mathbb{A}(\Sigma)$ in Definition~\ref{Def:ASHS}, we added an additional variable $z$ to the state tuple of the system $\Sigma$ as a counter that allows or prevents the system from jumping depending on its value. Since state trajectories of ASHSs and original SHSs are equivalent, we employ ASHSs in the sequel as a \emph{unified framework} covering both continuous evolutions and instantaneous jumps, which is more tractable to deal with.
\end{remark}

In the next section, we introduce a notion of augmented control barrier certificates for augmented SHSs. We then employ this notion and formally provide an upper bound on the probability that an augmented SHS reaches an unsafe region in a finite time horizon.

\section{Augmented Control Barrier Certificates}\label{sec:SPSF}

Augmented control barrier certificates in our work are defined as the following.

\begin{definition}\label{Def_1a} 
	Consider an augmented SHS $\mathbb{A}(\Sigma)=(\mathbb X,\mathbb U,\sigma,\rho,\varsigma, \mathbb F,\mathbb Y,\mathbb H)$. Let us define $\mathbb X_0=X_0\times \{0\},~\mathbb{X}_u = X_u \times \{0,\ldots,q_2\}$, as initial and unsafe sets of the augmented SHS, respectively, where $X_0, X_u \subseteq  \mathbb{R}^n$ are, respectively, initial and unsafe sets  of the original SHS $\Sigma$.
	A function $\mathcal B:\mathbb{X}\to\R_{\ge0}$ is called an augmented control barrier certificate (ACBC) for $\mathbb{A}(\Sigma)$ if there exist constants $0<\kappa<1$, $\alpha,\eta,\gamma\in\R_{\geq 0}$ with $\eta > \alpha$, such that
	\begin{align}\label{Eq_2a1}
	&\mathcal B(x,z) \leq \alpha,\quad\quad\quad\quad\quad\quad\quad\quad\!\!\! \forall(x,z)\in \mathbb{X}_0,\\\label{Eq_2a2}
	&\mathcal B(x,z) \geq \eta, \quad\quad\quad\quad\quad\quad\quad\quad\!\! \forall(x,z)\in \mathbb{X}_u, 
	\end{align}  
	and $\forall(x,z)\in \mathbb{X}$, $\exists \nu\in \mathbb U$, such that one has $(x',z')= \mathbb F((x,z),\nu,\sigma,\rho,\varsigma)$, and
	\begin{align}\label{Eq_3a}
	\EE \Big[\mathcal B(x',z')\,\big|\,x,\nu,z\Big]\leq \kappa \mathcal B(x,z)+\gamma,
	\end{align}
	where the expectation operator $\EE$ is with respect to $\varsigma$ under the one-step transition of the augmented SHS $\mathbb{A}(\Sigma)$.
\end{definition}

\begin{remark}
	Note that we need $\eta > \alpha$ in order to propose a meaningful probabilistic bound using Theorem~\ref{Kushner}. One can readily verify that the probabilistic safety guarantee in Theorem~\ref{Kushner} is improved by increasing the distance between initial- and unsafe-level sets of ACBC, \emph{i.e.,} $\alpha,\eta$.
\end{remark}

Now, by employing Definition~\ref{Def_1a}, we provide an upper bound on the probability that an augmented SHS reaches an unsafe region in a finite time horizon. The next theorem is borrowed from~\cite{1967stochastic} but adapted for augmented stochastic hybrid systems.

\begin{theorem}\label{Kushner}
	Let $\mathbb{A}(\Sigma)=(\mathbb X,\mathbb U,\sigma,\rho,\varsigma, \mathbb F,\mathbb Y,\mathbb H)$ be an augmented SHS. Suppose $\mathcal B$ is an ACBC for $\mathbb{A}(\Sigma)$ as in Definition~\ref{Def_1a}. Then for any random variable $x_0$ as the initial state and $z_0$ as the initial counter, the probability that the augmented SHS reaches an unsafe set $\mathbb X_u$ within the finite time horizon $k\in [0,\mathcal T]$ is upper bounded by $\delta$ as
	\begin{equation}\label{eqlemma2}
	\PP \Big\{\sup_{0 \leq k \leq \mathcal T} \mathcal B(x(t_k),z(t_k)) \geq \eta \,\, \big|\,\, x_0,z_0\Big\} \leq \delta,
	\end{equation}
	where	
	\begin{equation*}
	\delta=  \begin{cases} 
	1-(1-\frac{\alpha}{\eta})(1-\frac{\gamma}{\eta})^{\mathcal T}\!, & \quad\quad\quad\text{if } \eta \geq \frac{\gamma}{{1-\kappa}}, \\
	(\frac{\alpha}{\eta}){\kappa}^{\mathcal T}+(\frac{\gamma}{(1-{\kappa})\eta})(1-\kappa^{\mathcal T}), & \quad\quad\quad\text{if } \eta< \frac{\gamma}{{1-\kappa}}.  \\
	\end{cases}
	\end{equation*}	
\end{theorem}

In the next section, we provide required conditions for the construction of an ACBC for augmented SHSs.

\section{Construction of ACBC}\label{sec:constrcution_finite}

Here, in order to construct an ACBC for $\mathbb{A}(\Sigma)$, we first raise the following assumptions and lemma over original SHS $\Sigma$. As the first assumption, the SHS $\Sigma$ should have control barrier certificates (CBC) as in the following definition.

\begin{definition}\label{cbc}
	Consider an SHS $\Sigma$ and sets $X_0, X_u \subseteq X$ as its initial and unsafe sets, respectively. A function $\bar{\mathcal B}:X \rightarrow \mathbb{R}_{\geq 0}$ is said to be a control barrier certificate (CBC) for $\Sigma$ if there exist constants $\kappa_1\in\R,\kappa_2\in\R_{>0}$, $\gamma_1, \gamma_2,\bar\alpha,\bar\eta\in\R_{\geq 0}$, with $\bar\eta > \bar\alpha$, such that
	\begin{align}\label{subsys2}
	&\bar{\mathcal B}(x) \leq \bar\alpha,\quad\quad\quad\quad\quad\quad\!\!\forall x \in X_{0},\\\label{subsys3}
	&\bar{\mathcal B}(x) \geq \bar\eta, \quad\quad\quad\quad\quad\quad\!\forall x \in X_{u}, 
	\end{align} 
	\begin{itemize} 
		\item $\forall x\in \R^n$, $\exists \nu\in U$ such that,
	\end{itemize}
	\begin{align}\label{Eq:7}
	&\mathcal{L}\bar{\mathcal B}(x)\leq -\kappa_1 \bar{\mathcal B}(x) + \gamma_1,
	\end{align}
	with $\mathcal{L}\bar{\mathcal B}(x)$ being an infinitesimal generator of the stochastic process acting on the function $\bar{\mathcal B}(x)$ defined as~\cite{oksendal2013stochastic} 
	\begin{align}\label{Eq:8}
	\mathcal{L}\bar{\mathcal B}(x) &= \partial_x \bar{\mathcal B}(x) f_1(x,\nu)+ \frac{1}{2}\textsf{Tr}(\sigma(x)\sigma(x)^\top\partial_{x,x}\bar{\mathcal B}(x)) + \sum_{j=1}^{\textsf r}\lambda_j (\,\bar{\mathcal B}(x+\rho(x)\textsf e_j^\textsf r)- \bar{\mathcal B}(x)),
	\end{align}
	where  $\partial_x \bar{\mathcal B}(x) = \big [\frac{\partial \bar{\mathcal B}(x)}{\partial x_i}\big ]_{i}$ is a row vector, $\partial_{x,x}\bar{\mathcal B}(x) = \big [\frac{\partial^2{\bar{\mathcal B}(x)}}{\partial x_i \partial x_j}\big ]_{i,j}$, $\lambda_j$ is the rate of Poisson processes, and $\textsf e_j^\textsf r$ denotes an $\textsf r$-dimensional vector with $1$ on the $j$-th entry and $0$ elsewhere;
	\begin{itemize} 
		\item $\forall x\in \R^n$, $\exists \nu\in U$ such that,
	\end{itemize}
	\begin{align}\label{csbceq}
	\EE\Big[\bar{\mathcal B}(f_2(x,\nu)) \,\big|\, x, \nu\Big]\leq \kappa_2\bar{\mathcal B}(x) + \gamma_2.
	\end{align}
\end{definition}	
\vspace{0.2cm}
In addition to Definition~\ref{cbc}, we need to raise the following lemma to provide an upper bound on the evolution of the function $\bar{\mathcal B}$ which is required to show the main result of this section.

\begin{lemma}\label{Lemma:1}
	Consider a stochastic hybrid system $\Sigma=(\mathbb{R}^n, U,\mathcal{U}_\tau,\sigma,\rho,f_1,\varsigma,f_2)$ with jump parameters ($\tau$, $q_1$, $q_2$), where $\mathcal{U}_{\tau}$ is given according to \eqref{Eq:2}. Let \eqref{Eq:7} in Definition~\ref{cbc} hold. Then
	for all $x\in \R^n$, for all $\nu\in \mathcal{U}_{\tau}$, and for any two consecutive time instances $(t_{k},t_{k+1})$, one has  
	\begin{align}\label{Eq:11}
	\EE& \big[\bar{\mathcal B}(x_{x,\nu}(t_{k+1}^-))\,\big|\,x,\nu\big]\leq  e^{-\kappa_1(t_{k+1}-t_{k})}\big(\bar{\mathcal B}(x_{x,\nu}(t_{k})) + (t_{k+1} - t_k) \gamma_1\big).
	\end{align}
\end{lemma}

Under Definition~\ref{cbc}, as a sufficient condition, and Lemma~\ref{Lemma:1}, the next theorem lays the foundations for constructing an ACBC for $\mathbb{A}(\Sigma)$.

\begin{theorem}\label{Thm:Main}
	Consider a stochastic hybrid system $\Sigma=(\mathbb{R}^n, U,\mathcal{U}_\tau,\sigma,\rho,f_1,\varsigma,f_2)$ with its associated ASHS $\mathbb{A}(\Sigma)=(\mathbb X,\mathbb U,\sigma,\rho,\varsigma, \mathbb F,\mathbb Y,\mathbb H)$. Let $\bar{\mathcal B}$ be a CBC for  $\Sigma$, as in Definition~\ref{cbc}. If
	\begin{align}\label{Eq:122}
	\beta_{\bar\eta}\bar \eta > \beta_{\bar\alpha}\bar \alpha, 	\end{align}
	with
	\begin{align*}
	\beta_{\bar\eta}\Let\left\{\hspace{-0.5mm}
	\begin{array}{lr} 
	1,    \quad\quad\quad\quad\,~~~~ \text{if}~ \kappa_1>0~\&~0<\kappa_{2}<1,\\
	e^{\kappa_1\tau \epsilon_1 q_1},\quad\quad\!\! \text{if}~\kappa_1>0 ~\&~ \kappa_{2}\geq1,\\
	\kappa_{2}^{\frac{q_2}{\epsilon_2}},\quad\quad\quad\quad~ \text{if}~\kappa_1\leq0 ~\&~ 0<\kappa_{2}<1,
	\end{array}\right.
	\end{align*}
	\begin{align*}
	\beta_{\bar\alpha}\Let\left\{\hspace{-0.5mm}
	\begin{array}{lr} 
	1,    \quad\quad\quad\quad\,~~~~ \text{if}~ \kappa_1>0~\&~0<\kappa_{2}<1,\\
	e^{\kappa_1\tau \epsilon_1 q_2},\quad\quad\!\! \text{if}~\kappa_1>0 ~\&~ \kappa_{2}\geq1,\\
	\kappa_{2}^{\frac{q_1}{\epsilon_2}},\quad\quad\quad\quad~ \text{if}~\kappa_1\leq0 ~\&~ 0<\kappa_{2}<1,
	\end{array}\right.
	\end{align*}
	and
	\begin{align}\label{Eq:12}
	\ln(\kappa_{2})-\kappa_1\tau z<0, \quad \forall z\in\{q_1,\dots,q_2\},
	\end{align}
	then the function $\mathcal{B}$ defined as 
	\begin{align}\label{Eq:13}
	\mathcal{B}(x,z)=\beta~\!
	\bar{\mathcal B}(x)
	\end{align}
	is an ACBC for $\mathbb{A}(\Sigma)$ with
	\begin{align}\label{beta}
	\beta\Let\left\{\hspace{-0.5mm}
	\begin{array}{lr} 
	1,    \quad\quad\quad\,~~~~ \text{if}~ \kappa_1>0~\&~0<\kappa_{2}<1,\\
	e^{\kappa_1\tau \epsilon_1 z},\quad \text{if}~\kappa_1>0 ~\&~ \kappa_{2}\geq1,\\
	\kappa_{2}^{\frac{z}{\epsilon_2}},\quad\quad\quad~ \text{if}~\kappa_1\leq0 ~\&~ 0<\kappa_{2}<1,
	\end{array}\right.
	\end{align}
	for some $0<\epsilon_1<1$ and $\epsilon_2>q_2$. Accordingly, $\alpha = \beta_{\bar\alpha}\bar \alpha$, $\eta = \beta_{\bar\eta}\bar \eta$, and
	\begin{align*}
	\kappa\!\Let\!\left\{\hspace{-1.5mm}
	\begin{array}{lr} 
	\max\{e^{-\kappa_1\tau},\kappa_{2}\},    \quad\quad\quad\quad\quad\quad\quad\!\! \text{if}~ \kappa_1>0~\&~0<\kappa_{2}<1,\\
	\max\{e^{-\kappa_1\tau(1-\epsilon_1)},e^{-\kappa_1\tau \epsilon_1 q_1}\kappa_{2}\},\quad\!\!\! \text{if}~\kappa_1>0 ~\&~ \kappa_{2}\!\geq\! 1,\\
	\max\{e^{-\kappa_1\tau}\kappa_{2}^{\frac{1}{\epsilon_2}},\kappa_{2}^{\frac{\epsilon_2-q_2}{\epsilon_2}}\},\quad\quad\quad\quad\!\!\!\! \text{if}~\kappa_1\leq0 ~\&~ 0<\kappa_{2}<1,
	\end{array}\right.
	\end{align*}\vspace{-0.3cm}
	\begin{align*}
	\gamma\!\Let\!\left\{\hspace{-1.5mm}
	\begin{array}{lr} 
	\max\{e^{-\kappa_1\tau}\tau \gamma_1,\gamma_2\}, \quad\quad\quad\quad\quad\!\!\!\! \text{if}~ \kappa_1>0~\&~0<\kappa_{2}<1,\\
	\max\{e^{\kappa_1\tau \epsilon_1 q_2}e^{-\kappa_1\tau}\tau\gamma_1,\gamma_2\},\quad\quad~\! \text{if}~\kappa_1>0 ~\&~ \kappa_{2}\!\geq\!1,\\
	\max\{\kappa_{2}^{\frac{1}{\epsilon_2}}e^{-\kappa_1\tau}\tau \gamma_1,\gamma_2\},\quad\quad\quad\quad~~\!\!\! \text{if}~\kappa_1\leq0 ~\&~ 0<\kappa_{2}<1.
	\end{array}\right.
	\end{align*}
	
\end{theorem}

\section{Computation of CBC}\label{compute_CBC}
In  this section, we provide a systematic approach to search for CBC and its corresponding controllers. The employed approach is based on sum-of-squares (SOS) optimization~\cite{parrilo2003semidefinite} in which one
can reformulate conditions~\eqref{subsys2}-\eqref{Eq:7}, \eqref{csbceq} as an SOS optimization problem, where the CBC is restricted to be
non-negative as a sum of squares
of different polynomials. In order to utilize the SOS optimization, the following assumption is essential.

\begin{assumption}\label{ass:BC}
	Assume that $\Sigma$ has a continuous state set $X\subseteq \mathbb R^{n}$ and a continuous input set $U\subseteq \R^{m}$. Moreover, $f_1, f_2, \sigma, \rho$ are all polynomial functions.
\end{assumption}

Under Assumption~\ref{ass:BC}, the following lemma provides the SOS formulation.

\begin{lemma}\label{sos}
	Suppose Assumption~\ref{ass:BC} holds and sets $X_0,X_u, X, U$ can be defined by vectors of polynomial inequalities as $X_{0}=\{x\in\R^{n}\mid g_{0}(x)\geq0\}$, $X_{u}=\{x\in\R^{n}\mid g_{u}(x)\geq0\}$, $X=\{x\in\R^{n}\mid g(x)\geq0\}$, and $U=\{U\in\R^{m}\mid g_{\nu}(\nu)\geq0\}$, where the inequalities are defined element-wise.
	Suppose there exists an SOS polynomial $\bar{\mathcal B}(x)$, constants $\kappa_1\in\R,\kappa_2\in\R_{>0}$, $\gamma_1,\gamma_2,\bar\alpha,\bar\eta\in\R_{\geq 0}$, polynomials $l_{\nu_{j}}(x)$, $\bar l_{\nu_{j}}(x)$ corresponding to the $j^{\text{th}}$ input in $\nu=(\nu_{1},\nu_{2},\ldots,\nu_{m})\in U\subseteq \R^{m}$ and $\bar\nu=(\bar\nu_{1},\bar\nu_{2},\ldots,\bar\nu_{m})\in U\subseteq \R^{m}$, respectively, and vectors of sum-of-squares polynomials $l_{0}(x)$, $l_{u}(x)$, $l(x,\nu), \hat l(x,\nu), l_\nu(x,\nu)$, and $\hat l_\nu(x,\nu)$, of appropriate dimensions such that the following expressions are sum-of-squares polynomials:
	\begin{align}\label{eq:sos1}
	-&\bar{\mathcal B}(x)-l_{0}^\top(x) g_{0}(x)+\bar\alpha,\\\label{eq:sos2}
	&\bar{\mathcal B}(x)-l_{u}^\top(x) g_{u}(x)-\bar\eta,\\\label{eq:sos3}
	-&\mathcal{L}\bar{\mathcal B}(x) - \kappa_1 (\bar{\mathcal B}(x)) +\gamma_1
	-\sum_{j=1}^{m}(\nu_{j}-l_{\nu_{j}}(x))- l^\top(x,\nu) g(x)-l_{\nu}^\top(x,\nu) g_{\nu}(\nu),\\\label{eq:sos4}
	-& \EE\Big[\bar{\mathcal B}(f_2(x,\nu)) \,\big|\, x, \nu\Big] + \kappa_2\bar{\mathcal B}(x) + \gamma_2 - \sum_{j=1}^{m}(\bar\nu_{j}-\bar l_{\nu_{j}}(x)) - \hat l^\top(x,\nu) g(x) - \hat l_{\nu}^\top(x,\nu) g_{\nu}(\nu).
	\end{align}
	Then $\bar{\mathcal B}(x)$ satisfies conditions~\eqref{subsys2}-\eqref{Eq:7}, \eqref{csbceq} in Definition~\ref{cbc}. In addition, $\nu=[l_{\nu_{1}}(x);\dots;l_{\nu_{m}}(x)]$ and $\bar\nu=[\bar l_{\nu_{1}}(x);\dots;\bar l_{\nu_{m}}(x)]$ are corresponding controllers in flow and jump scenarios, respectively.
\end{lemma} 

\section{Case Study}\label{Sec:Case}
We demonstrate the effectiveness of our proposed results by applying them to a nonlinear SHS $\Sigma$ as 
\begin{align}\label{Eq:14}
\Sigma\!:\left\{\hspace{-2mm}
\begin{array}{rl}
\mathsf{d}x(t)\!=\!\!\!\!& (a_1x^3(t)+b_1\nu(t))\mathsf{d}t + 0.6\mathsf{d}\mathbb W_t+0.5\mathsf{d}\mathbb P_t,
~ t\in\mathbb{R}_{\geq0}\backslash \Lambda,\\
x(t)\!=\!\!\!\!& a_2x^3(t^-)+ b_2\nu(t)+0.5\varsigma(t), \quad\quad\quad\quad\quad\quad\quad ~~~\!\!\!\!t\in \Lambda,
\end{array}
\right.
\end{align}
with $a_1,a_2,b_1,b_2 \in \mathbb R$, jump parameters $\tau= 0.1$, $q_1 = 1,q_2 = 7$, and the rate of Poisson processes as $\lambda = 0.5$. In addition, the regions of interest are given as $X \in [0,8], X_{0} \in [0,1.5], X_{u} = [7,8]$. The main goal is to design an ACBC for the augmented system and its corresponding safety controllers such that the state of the system remains in the comfort zone $[0,7]$. To do so, we first search for a CBC and accordingly design controllers for $\Sigma$. Consequently, an ACBC can be constructed from the CBC according to Theorem~\ref{Thm:Main}. We show our results for different values of $a_1,a_2,b_1,b_2$.

We employ the software tool \textsf{SOSTOOLS}~\cite{papachristodoulou2013sostools} and the SDP solver \textsf{SeDuMi}~\cite{sturm1999using} to compute CBC as described in Section~\ref{compute_CBC}. For $a_1 = -0.4,b_1 = 0.5,a_2 = 0.01,b_2 = 0.06$, based on Lemma~\ref{sos}, we compute a CBC of an order $4$ as $\bar{\mathcal B}(x) = 0.0054x^4 - 0.0345x^3 + 0.0814x^2 - 0.0849 x + 0.0369$ and corresponding controllers $\nu = -0.05152x + 3$ and $\bar\nu = -0.06145x + 2.6$ for flow and jump scenarios, respectively. Moreover, the corresponding constants in Definition~\ref{cbc} satisfying conditions~\eqref{subsys2}-\eqref{Eq:7}, \eqref{csbceq} are quantified as $\bar\alpha = 0.13, \bar\eta = 4.4,\gamma_1 = 0.0015, \gamma_2 = 0.0012$, $\kappa_{1} = 0.01, \kappa_2 = 0.99$.
We now proceed with Theorem~\ref{Thm:Main} to construct an  ACBC for the augmented system using the obtained CBC. We select $\epsilon_1 = 0.1, \epsilon_2 = 8$. Since  $\kappa_1 =  0.01>0$ and $0<\kappa_{2} = 0.99 <1$, the results of bound 1 are useful. Then condition~\eqref{Eq:122} is satisfied with $\bar\eta > \bar\alpha$. One can readily verify that condition~\eqref{Eq:12} is also met for all $z \in \{q_1,\dots,q_2\}$. Then one can conclude that $\mathcal B(x,z)\Let\bar{\mathcal B}(x)$ is a CBC for the augmented system with $\kappa=\max\{e^{-\kappa_1\tau},\kappa_{2}\} = 0.99$ and $\gamma=\max\{e^{-\kappa_1\tau}\tau \gamma_1,\gamma_2\} = 0.0012$.

For $a_1 = -0.3,b_1 = 0.2,a_2 = 1.01,b_2 = 1$, we compute a CBC of an order $4$ as $\bar{\mathcal B}(x) = 0.0061x^4 - 0.0438x^3 + 0.1163x^2 - 0.1375 x + 0.0617$ and corresponding controllers $\nu = -0.02152x + 4$ (flow scenario) and $\bar\nu = -0.99x + 2$ (jump scenario). Furthermore, the corresponding constants satisfying conditions~\eqref{subsys2}-\eqref{Eq:7}, \eqref{csbceq} are obtained as $\bar\alpha = 0.12, \bar\eta = 4.6,\gamma_1 = 0.0025, \gamma_2 = 0.003$, $\kappa_{1} = 0.04547, \kappa_2 = 1.00001$. Since  $\kappa_1 =  0.04547>0$ and $\kappa_{2} = 1.00001 \geq1$, bound 2 is valid. Then condition~\eqref{Eq:122} is satisfied with $1.0005\bar\eta > 1.0032\bar\alpha$. Condition~\eqref{Eq:12} is also met for all $z \in \{q_1,\dots,q_2\}$. Then $\mathcal B(x,z)\Let e^{4.54 z}\bar{\mathcal B}(x)$ is a CBC for the augmented system with $\kappa=\max\{e^{-\kappa_1\tau(1-\epsilon_1)},e^{-\kappa_1\tau \epsilon_1 q_1}\kappa_{2}\} = 0.99$ and $\gamma=\max\{e^{\kappa_1\tau \epsilon_1 q_2}e^{-\kappa_1\tau}\tau\gamma_1,\gamma_2\} = 0.003$. 

For $a_1 = 0.01,b_1 = 0.7,a_2 = 0.02,b_2 = 0.9$, we compute a CBC of an order $4$ as $\bar{\mathcal B}(x) = 0.0077x^4 - 0.0673x^3 + 0.2158x^2 - 0.3031 x + 0.1581$ and its corresponding controllers $\nu = -0.2852x + 2.5$ (flow scenario) and $\bar\nu = -0.19x + 3$ (jump scenario). In addition, the corresponding constants in Definition~\ref{cbc} are acquired as $\bar\alpha = 0.16, \bar\eta = 4.2,\gamma_1 = 0.003, \gamma_2 = 0.003$ and $\kappa_{1} = -0.0005, \kappa_2 = 0.98$. Since  $\kappa_1 = -0.0005\leq0$ and $0<\kappa_{2} = 0.98 <1$, bound 3 is useful. Then condition~\eqref{Eq:122} is satisfied with $0.9825 \bar\eta > 0.9975 \bar\alpha$. Condition~\eqref{Eq:12} is also met for all $z \in \{q_1,\dots,q_2\}$. Consequently, one can conclude that $\mathcal B(x,z)\Let 0.98^{\frac{z}{8}}\bar{\mathcal B}(x)$ is a CBC for the augmented system with $\kappa=\max\{e^{-\kappa_1\tau}\kappa_{2}^{\frac{1}{\epsilon_2}},\kappa_{2}^{\frac{\epsilon_2-q_2}{\epsilon_2}}\} = 0.997$ and $\gamma=\max\{\kappa_{2}^{\frac{1}{\epsilon_2}}e^{-\kappa_1\tau}\tau \gamma_1,\gamma_2\} = 0.003$.

By employing Theorem~\ref{Kushner}, we guarantee that the state trajectory of $\Sigma$ starting from initial conditions inside $X_{0} = [0~1.5]$ remains in the safe set $[0~7]$ during the time steps $\mathcal T=100$ (equivalent to 10 seconds) with the probability of at least $0.9443$ (first bound),  $0.9124$ (second bound), and $0.8939$ (last bound).

\begin{figure}
	\centering
	\includegraphics[width=5.5cm]{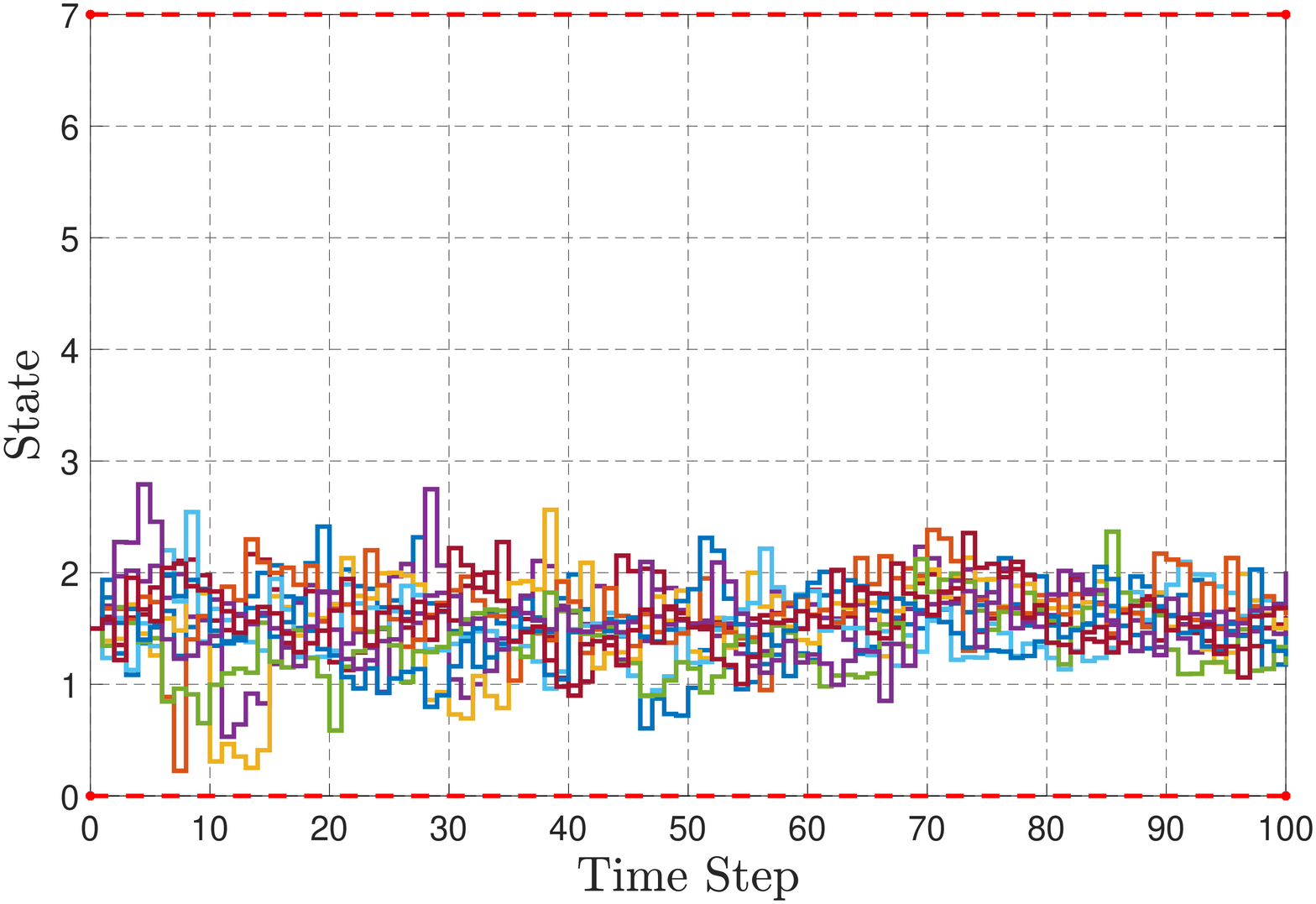}
	\includegraphics[width=5.5cm]{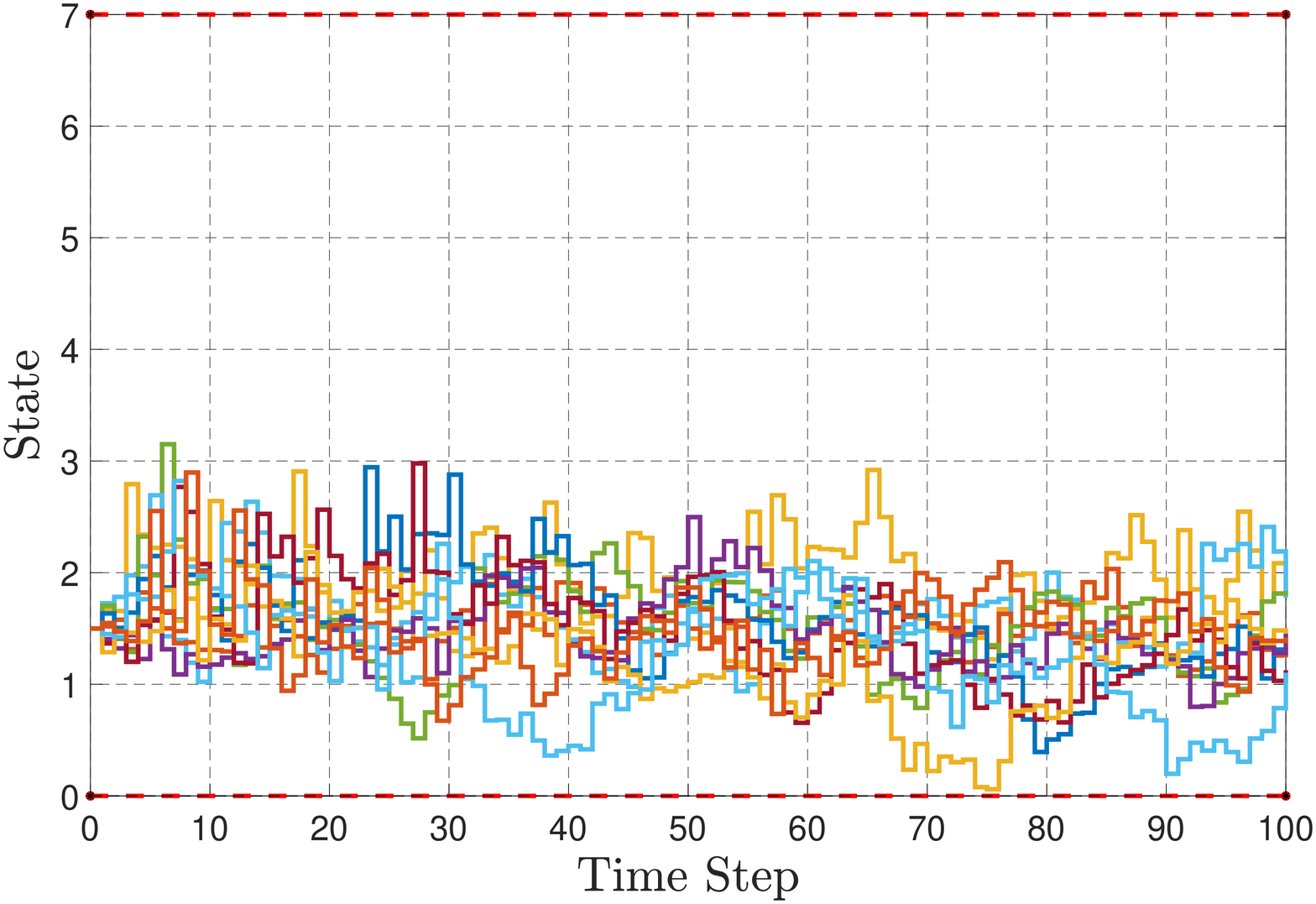}
	\includegraphics[width=5.5cm]{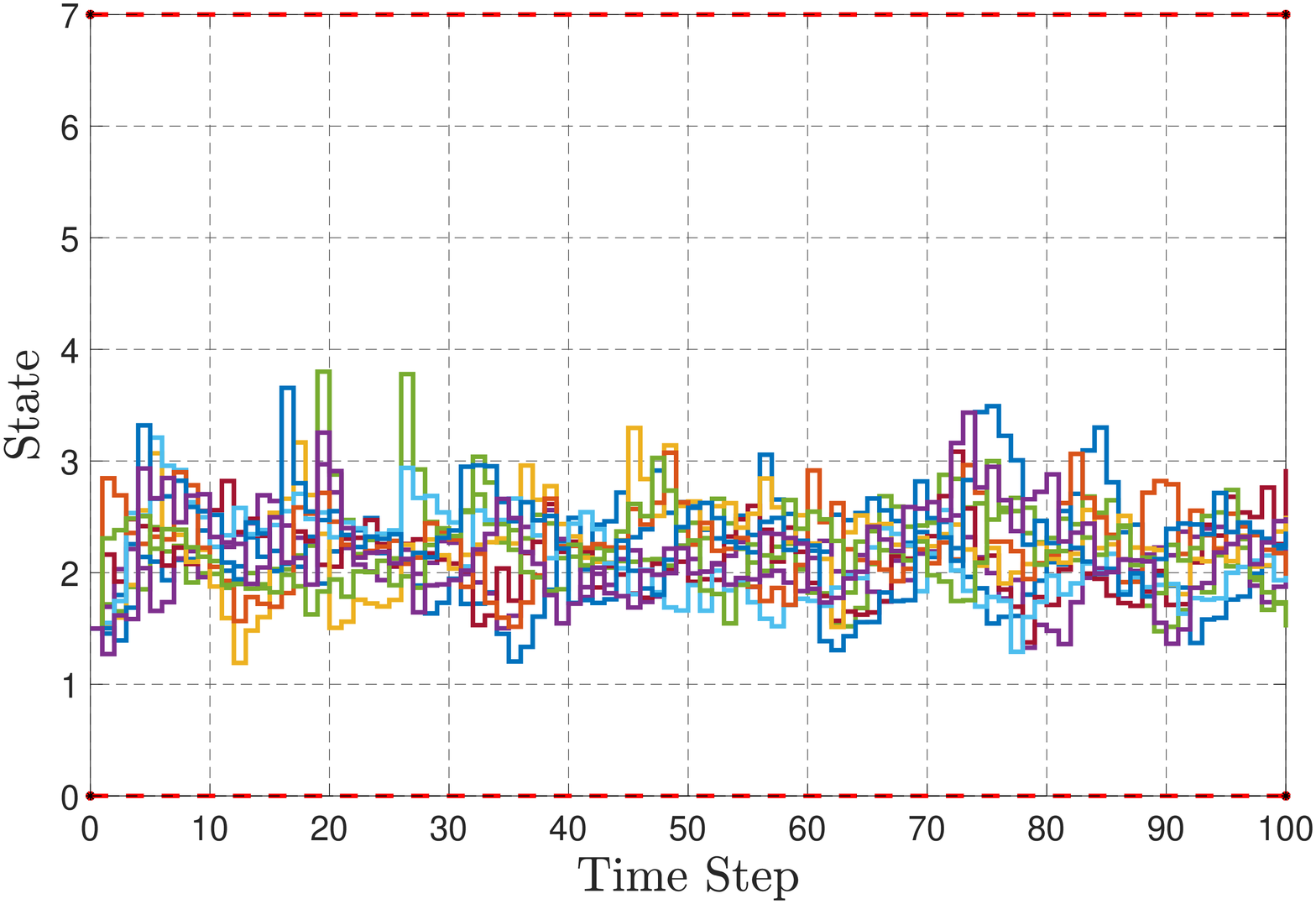}
	\caption{Closed-loop state trajectories of SHS $\Sigma$  with $10$ different noise realizations for  $a_1 = -0.4,b_1 = 0.5,a_2 = 0.01,b_2 = 0.06$ (left), $a_1 = -0.3,b_1 = 0.2,a_2 = 1.01,b_2 = 1$ (middle), $a_1 = 0.01,b_1 = 0.7,a_2 = 0.02,b_2 = 0.9$ (right).}
	\label{Simulation}
\end{figure}

Closed-loop state trajectories of SHS $\Sigma$ for different bounds with $10$ different noise realizations are depicted in Fig.~\ref{Simulation}. We have only plotted trajectories at the discrete time steps since our guarantee is provided for them. It is worth mentioning that by employing our synthesized controllers and running Monte Carlo simulations for the closed-loop system, the empirical probabilities are better than the ones we provided here. This issue is expected and the reason is due to the conservatism nature of polynomial barrier certificates that we employed here with a fixed degree, but with the gain of providing a formal lower bound on the safety specification rather than an empirical one. One solution for mitigating the aforementioned conservatism is to design higher-degree polynomials for barrier certificates and safety controllers. The computation of CBC and its corresponding controllers for each bound took almost $30$ seconds with a memory usage of $2.6$ MB on a Windows operating system (Intel i7@3.6GHz CPU and 32 GB of RAM).

\section{Conclusion}\label{Conclude}
In this work, we proposed a safety controller synthesis approach for continuous-space stochastic hybrid systems (SHSs): continuous evolutions are handled by stochastic differential equations with both Brownian motions and Poisson processes, and instantaneous jumps are governed by stochastic difference equations with additive noises. We employed \emph{control barrier certificates} (CBC) and synthesized safety controllers for SHSs while providing safety guarantees in finite time horizons. To do so, we first defined an augmented framework to describe each SHS containing continuous evolutions and instantaneous jumps via a single system covering both scenarios. We then introduced an augmented control barrier certificate (ACBC) for augmented SHS and provided required conditions to construct an ACBC based on the CBC of original hybrid systems. We leveraged the constructed ACBC and quantified upper bounds on the probability that the SHS reaches certain unsafe regions in a finite time horizon. We finally verified our results over a nonlinear case study. Providing a \emph{compositional approach} for the controller synthesis of \emph{large-scale interconnected SHSs} using the proposed techniques here is under investigation as a future work.

\bibliographystyle{IEEEtran}
\bibliography{biblio}

\end{document}